# An *ab–initio* study on physical properties of $Pd^{2+}$ incorporated double perovskites $CaPd_3B_4O_{12}$ (*B* = Ti, V)

**Mirza H. K. Rubel[1*], Khandaker Monower Hossain[1], Anjuman Ara Khatun[1], M. Anwar Hossain[2], M. M. Rahaman[1,3*], M. Mozahar Ali[4], M. M. Hossain[5], J. Hossain[6], Md. Rasadujjaman[7], S. Kojima[8], N. Kumada[9]**

[1]Department of Materials Science and Engineering, University of Rajshahi, Rajshahi–6205, Bangladesh.
[2]Department of Physics, Mawlana Bhashani Science and Technology University, Santosh, Tangail–1902, Bangladesh.
[3]Geophysical Laboratory, Carnegie Institution for Science, Washington, DC 20015, USA.
[4]Department of Physics, American International University–Bangladesh (AIUB), Ka–66/1, Kuratoli Road, Kuril, Khilkhet, Dhaka 1229, Bangladesh.
[5]Department of Physics, Chittagong University of Engineering and Technology, Chittagong–4349, Bangladesh.
[6]Department of Electrical and Electronic Engineering, University of Rajshahi, Rajshahi–6205, Bangladesh.
[7]Department of Physics, Dhaka University of Engineering & Technology, Gazipur-1707, Bangladesh.
[8]Graduate School of Pure and Applied Sciences, University of Tsukuba, Tsukuba, Ibaraki 305–8573, Japan.
[9]Center for Crystal Science and Technology, University of Yamanashi, 7–32 Miyamae–Cho, Kofu 400–8511, Japan.

*****Corresponding author:** Mirza H. K. Rubel; Md. Mijanur Rahaman
Cell: +8801714657365; E–mail: mhk_mse@ru.ac.bd; mijan_mse@ru.ac.bd

**Abstract**

Numerous physical properties of $CaPd_3Ti_4O_{12}$ (CPTO) and $CaPd_3V_4O_{12}$ (CPVO) double perovskites have been explored based on density functional theory (DFT). The calculated structural parameters fairly agree with the experimental data to confirm their stability. The mechanical stability of these two compounds was clearly observed by the Born stability criteria. To rationalize the mechanical behavior, we investigate elastic constants, bulk, shear and Young's modulus, Pugh's ratio, Poisson's ratio and elastic anisotropy index. The ductility index confirms that both materials are ductile in nature. The electronic band structure of CPTO and CPVO reveals the direct band gap semiconducting in nature and metallic characteristics, respectively. The calculated partial density of states indicates the strong hybridization between Pd $4d$ and O $2p$ orbital electrons for CPTO and Pd $4d$ and V $3d$ O $2p$ for CPVO. The study of electronic charge density map confirms the coexistence of covalent, ionic and metallic bonding for both compounds. Fermi surface calculation of CPVO ensures both electron and hole like surfaces indicating the multiple band nature. In the midst of optical properties, photoconductivity and absorption coefficient of both compounds reveal well qualitative compliance with consequences of band structure computations. Among the thermodynamic properties, the Debye temperature has been calculated to correlate its topical features including thermoelectric behavior. The studied thermoelectric transport properties of CPTO yielded the Seebeck coefficient (186 μV/K), power factor (11.9 μWcm⁻¹K⁻²) and figure of merit (ZT) value of about 0.8 at 800 K indicate that this material could be a promising candidate for thermoelectric device application.

**Key–words:** DFT calculations; Elastic properties; Electronic properties; Optical properties; Thermodynamic properties; Thermoelectric properties;

## 1. Introduction

Perovskite oxides with chemical formula $ABO_3$ and its various forms or modes are extensively studied owing to their diverse intriguing physical and chemical properties. The primarily induced properties are resulted due to interaction between the transition metal cations at the B site and/or B–O–B interactions through oxygen ions. A large variety of chemical substitution is possible in perovskite compounds at both A and B cations site. Double perovskite materials are one of these derivatives which have the general formula $A_2BB′O_6$ or $AA′BB′O_6$ where A and A′ are alkaline–earth and/or rare–earth metals and B and B′ are transition metals. Now–a–days double perovskites have been widely explored because of their numerous interesting physical and electrochemical properties [1–6]. These types of perovskites are also tremendously used in the design and technology of magnetic memories, nonlinear optics, magneto-optic, magneto-ferroics, thermoelectric properties, tunnel junctions and other magnetic devices in the novel spintronics arena [7-9]. An unusual ordered arrangement of the A–site ions results an A–site ordered perovskite–structure oxide with the formula $AA′_3B_4O_{12}$ where the A and A′ ions are in ordered state

[10]. Accordingly, among the A–site ordered perovskite novel compounds CaPd$_3$Ti$_4$O$_{12}$ (CPTO) and CaPd$_3$V$_4$O$_{12}$ (CPVO), in which the A'–sites are fully occupied by rare Pd$^{2+}$ ions [11]. In this type of perovskites, three–quarters of the A–sites (A'–site) have a pseudo square planar coordination with Jahn–Teller active ions such as Cu$^{2+}$ and Mn$^{2+}$, whereas the other remaining quarter of A–positions are filled by ideal large A–site ions like alkaline, alkaline earth, and rare earth metal ions [12−14]. In these structures transition–metal cations are set as twelve–fold coordinated A'–site form square–coordinated units which are normal to each other. This aligning of the A'O$_4$ squares along with the large tilting of the corner–sharing BO$_6$ octahedra make the structural 2a×2a ×2a unit cell that deviates from a simple perovskite structure. As these transition metal ions are positioned at the A'–site and the B–site, thus A'–A' and A'–B interactions and further B–B interaction are observed in ABO$_3$ perovskite–type oxides. In recent years such interactions have drawn significant attention due to their potential and useful properties [15, 16]. However, currently Rubel et al. [17] and several researchers have explored Bi–based superconductive AA'$_3$B$_4$O$_{12}$–type perovskite structure [17-19]. Furthermore, some theoretical studies have been implemented extensively on these Bi–oxide double and simple perovskites [20-22].

Recently, Kentaro Shiro et al. [11] have synthesized AA'$_3$B$_4$O$_{12}$–type two novel ordered perovskites CaPd$_3$Ti$_4$O$_{12}$ and CaPd$_3$V$_4$O$_{12}$ by entirely incorporating Pd$^{2+}$ ions in A'–site under high–pressure, high–temperature conditions (15 GPa and 1000 °C). The valence states of these perovskites Ca$^{2+}$Pd$^{2+}_3$Ti$^{4+}_4$O$_{12}$ and Ca$^{2+}$Pd$^{2+}_3$V$^{4+}_4$O$_{12}$ are confirmed by spectroscopic investigations. The authors also have reported the primal dependence of the metal–oxygen angles upon the difference between the A'– and A–site ion sizes for AA'$_3$B$_4$O$_{12}$–type perovskite structure. Moreover, the measurements of magnetic property, electrical resistivity and specific heat ensured that the double perovskite CPTO is a diamagnetic insulator while CaPd$_3$V$_4$O$_{12}$ was treated as Pauli–paramagnetic metal. Nonetheless, no research work has been carried out to pursue additional physical properties of CPTO and CPVO double perovskites for device applications. Therefore, these two crystalline double perovskites is our point of interest to investigate unexplored physical features by using the density functional theory (DFT). In this paper, the calculated structural, elastic, electronic, and bonding properties of the aforementioned compounds are discussed. However, the authors did not compute or publish any report on other physical properties such as band structure, elastic, optical, thermodynamic or thermoelectric properties of CPVO and CPTO to examine their potentiality for device applications. Among these physical peculiarities, thermoelectric properties is utilized for the thermoelectric power generation which might be regarded as a long-term environment friendly energy harvesting technology that can convert the thermal energy directly into electrical energy and vice-versa with superior durability making no noise or vibration. Since, thermoelectric conversion technology has drawn significant attraction, thus perovskite oxides with thermoelectric properties can potentially replace conventional thermoelectric materials because of their unique and comparatively simple form of structure. Now a days, the thermoelectric performance of complex perovskite oxides for instance Sr$_{2-x}$M$_x$B'MoO$_6$: M = Ba, La, B' = Fe, Mn [23-27] have been studied from room to elevated temperature by several researchers those exhibited relatively better efficiency than that of conventional oxides. However, the conversion efficiency of a thermoelectric compound is evaluated in terms of its dimensionless figure of merit, Z = (S$^2$σ/K)T, where the parameters S, σ, K and T denote the Seebeck coefficient, electrical conductivity, thermal conductivity and absolute temperature, respectively. However, these perovskites are yet to employ practically with ZT < 1.0 condition. The band structure calculation of CPTO showed a direct band gap, is a property of semiconductor which inspired us to perform thermoelectric properties of this material only. Moreover, the narrow band semiconductors which are made of p- and n-type carriers are considered as promising candidates for thermoelectric device applications [20, 28]. To the best of our knowledge, there is no report of temperature dependent transport properties on the basis of theoretical and experimental methodology for CPTO double perovskite. Therefore, the aforementioned scenarios have stimulated us to study various unexplored physical properties of A–site ordered CaPd$_3$Ti$_4$O$_{12}$ and CaPd$_3$V$_4$O$_{12}$ double perovskites based on DFT calculations by employing CASTEP and WIEN2k programs. Consequently, we calculate and compare the structural, mechanical, electronic (band structure, DOS, charge density map and Fermi surface), optical, population analysis, thermodynamic and thermoelectric properties of recently synthesized CaPd$_3$B$_4$O$_{12}$ (B = Ti, V) double perovskites.

## 2. Computational method

The present calculations are carried out by employing the Cambridge Serial Total Energy Package (CASTEP) [29, 30] and WIEN2k [31] programs based on the first–principles density functional theory (DFT) [32, 33]. The generalized gradient approximation (GGA) of Perdew–Burke–Ernzerhoffor solid (PBEsol) [34] is utilized to evaluate the electronic exchange and correlation potentials. Vanderbilt–type ultrasoft pseudo potential [35] is also used to represent the electrostatic interaction between valence electron and ionic core. The cutoff energy for the plane wave expansion is chosen as 700 eV and a *k*–point mesh of 6×6×6 is used for integration over the first

Brillouin zone according to Monkhorst–Pack [36, 37] scheme. We also apply the Broyden–Fletcher–Goldfarb–Shanno (BFGS) algorithm [38] to optimize the atomic configuration as well as density mixing is employed to optimize the electronic structure. Convergence tolerance for energy, maximum force, maximum displacement, and maximum stress are chosen as $2\times10^{-5}$ eV/atom, 0.05 eV/Å, 0.02 Å and 0.1 GPa, respectively in the geometry optimization. The thermodynamic properties of the materials are calculated using IRelast program interfaced with Wien2k code [39]. Various thermoelectric properties have been computed through resolving Boltzmann semi-classical transport equations as employed in BoltzTrap [40] interfaced with WIEN2k [31] program. A plane wave cut-off of kinetic energy $RK_{max}$ = 7.0 was chosen to produce a good convergence for the self-consistent field (SCF) calculations. A mesh of 21×21×21 k-points was used for thermoelectric properties calculation. As GGA-PBE potential [41, 42] underestimates the band gap of semiconductors, thus Tran-Blaha modified Becke-Johnson potential (TB-mBJ) [43] was also used in the calculation of electronic, transport and optical properties. The muffin tin radii 2.46, 2.09, 1.94 and 1.75 Bohr, respectively for Ca, Pd, Ti and O were fixed in the calculations. During the calculation of temperature dependent transport properties the value of chemical potential was fixed to that of Fermi energy. The parameter relaxation time ($\tau$) was considered as a constant. The electronic conductivity as well as the electronic part of thermal conductivity was accounted in terms of relaxation time, on the other hand the Seebeck coefficient was independent of it.

## 3. Results and discussion

### 3.1. Structural Properties

The two quaternary $CaPd_3Ti_4O_{12}$ and $CaPd_3V_4O_{12}$ double perovskites crystallize in a cubic structure with the space group of $Im\bar{3}$ (No. 204). The equilibrium lattice parameters of $CaPd_3Ti_4O_{12}$ and $CaPd_3V_4O_{12}$ compounds are 7.49777 and 7.40317, respectively [11].

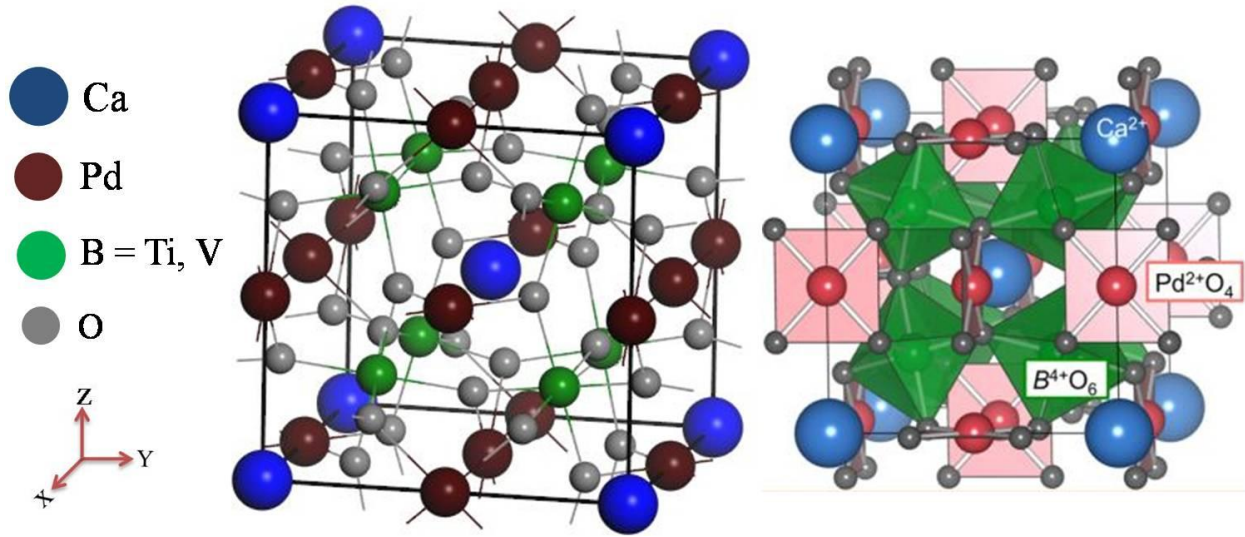

**Figure 1.** (Color online) Unit cell of cubic ($Im\bar{3}$) $CaPd_3B_4O_{12}$ ($B$ = Ti, V) perovskites (left−side) and three types of octahedra in the unit cell are shown the blue (A), red (A′), green (B), and gray (O) spheres represent the relevant atoms. The unit cell is built up by the atoms inside the solid lines: eight distorted $BO_6$ octahedra relative to one another, A′ ions bonded to four closest oxygen atoms and A ions at the corners as well as at the body–center (right−side) (reproduced from Ref. [11]).

In the optimized structures, the atomic position of Ca, Pd and Ti/V are (0, 0, 0), (0, 0.5, 0.5) and (0.25, 0.25, 0.25), respectively. Besides, the atomic positions of O are (0.2961, 0.1859, 0) and (0.2947, 0.1856, 0) respectively for $CaPd_3Ti_4O_{12}$ and $CaPd_3V_4O_{12}$. The crystal structure of these compounds as a structural model of cubic $CaPd_3B_4O_{12}$ ($B$ = Ti, V) system is depicted in Fig. 1. Table 1 represents the unit cell parameters of titled $CaPd_3B_4O_{12}$ perovskites. The calculated lattice constants are well consistent with the available experimental values [11].

**Table 1.** Calculated lattice constants ($a$) in Å and Volume ($V$ in Å$^3$) of CaPd$_3B_4$O$_{12}$ ($B$ = Ti, V) compounds.

| Compounds | Lattice constant, $a$ (Experimental) | Lattice constant, $a$ (Present study) | Volume, $V$ (Present study) |
|---|---|---|---|
| CaPd$_3$Ti$_4$O$_{12}$ | 7.49777 (14) [11] | 7.593643 | 415.552490 |
| CaPd$_3$V$_4$O$_{12}$ | 7.40317(8) [11] | 7.392642 | 395.270504 |

### 3.2. Mechanical Properties

The elastic constants of a material provide extensive understanding about the response of the material to external stresses within the elastic limit. The nature of bonding in solids is related to their mechanical properties. To show mechanical stability, the cubic crystal should fulfill the Born criteria [44]: $C_{11}+ 2C_{12}> 0$, $C_{11}> 0$, $C_{44}> 0$, $C_{11}− C_{12}> 0$. The calculated three independent elastic constants, shown in Table 2, which completely satisfy the above conditions, indicating that the material CaPd$_3B_4$O$_{12}$ ($B$ = Ti, V) are mechanically stable.

The polycrystalline elastic constants such as the bulk modulus ($B$), Young's modulus (Y), shear modulus ($G$) and Poisson's ratio ($v$) are calculated from the single crystal zero–pressure elastic constants using the Voigt($V$)–Reuss($R$)–Hill($H$) formula [44–47]. All the calculated elastic parameters are listed in Table 2. The bulk modulus evaluates the average bond strength of constituent atoms for a given material [48]. The calculated value of $B$ (243 GPa and 242 GPa for CaPd$_3$Ti$_4$O$_{12}$ and CaPd$_3$V$_4$O$_{12}$, respectively) confirm the strong bonding strength of atoms involved in these compounds. The bond strength of atoms also offers the required resistance to volume deformation under the action of external pressure. On the contrary, change of shape in a solid largely depends on its shear modulus $G$, which shows a crucial correlation with material's hardness. To estimate the material's hardness directly, Vickers hardness parameter is frequently used and defined by: $Hv = 2(k^2G)^{0.585}−3$ [49] where, $k$ (= $G/B$) is the Pugh's ratio. The calculated values of the Pugh's ratio ($k$), and Vickers hardness ($Hv$), Poisson's ratio ($v$), and Cauchy pressure are given in Table 2. It is established that Diamond is the hardest material which has Vickers hardness in the range of 70 to 150 GPa. It is seen from Table 2 that the calculated values of the Vickers hardness for CPTO and CPVO are 7.96 GPa and 6.60 GPa, respectively. Therefore, we can conclude that the compounds under this study are relatively very soft in contrast to diamond.

For most of the practical situations, a material is needed to identify as either ductile or brittle. A material is said to be ductile if the value of Pugh's ratio is smaller than 0.57 [50]; otherwise, it is brittle. From Table 2, the values of Pugh's ratio are 0.43 and 0.39, respectively for CaPd$_3$Ti$_4$O$_{12}$ and CaPd$_3$V$_4$O$_{12}$, indicating the ductile nature of CaPd$_3B_4$O$_{12}$. Frantsevich's also proposed a critical value of Poisson's ratio ($v \geq 0.26$) for understanding the ductile nature of solids [51]. The values of calculated Poisson's ratio ($v$) for CaPd$_3$Ti$_4$O$_{12}$ and CaPd$_3$V$_4$O$_{12}$ are 0.31 and 0.32 [Table 2], also predict the ductile nature of these compounds. Furthermore, another indicator to identify ductile/brittle nature of a material is the Cauchy pressure which is expressed as ($C_{12}−C_{44}$) [52]. If the Cauchy pressure is negative, the material is expected to be brittle; otherwise (having positive Cauchy pressure), it demonstrates the ductile behavior [52]. Since, the Cauchy pressure is positive for both compounds, they possess ductile nature. Therefore, the compounds CaPd$_3B_4$O$_{12}$ in this study show ductile characteristics in accordance with the above mentioned three indicators.

The universal anisotropic index, denoted by $A^U = 5\frac{G_V}{G_R} + \frac{B_V}{B_R} − 6 \geq 0$, is an indicator of crystal anisotropy has been used to calculate the anisotropic nature of the materials. When $A^U = 0$, the material is perfectly isotropic but the deviation from zero reveals the extent of elastic anisotropy of crystals [53]. The values of $A^U$ for CaPd$_3$Ti$_4$O$_{12}$ and CaPd$_3$V$_4$O$_{12}$ are 0.15 and 0.19 [Table 2], respectively implying significant anisotropy of these materials.

In this study, we also have calculated Peierls stress ($\sigma_p$) to predict the strength of CaPd$_3B_4$O$_{12}$ for motile a dislocation inside the atomic plane by using the expression [54] as follows:

$$\sigma_p = \frac{G}{1-v} \exp\left[-\frac{2\pi d}{b(1-v)}\right]$$

Herein, $G$, $v$, $b$ and $d$ are the shear modulus, Poisson ratio, Burgers vector and interlayer distance between the glide planes (Table 2). The calculated $\sigma_p$ of CaPd$_3B_4$O$_{12}$ is found to be 1.51 and 1.39, respectively for CPTO and CPVO. These values of $\sigma_p$ of double perovskites CaPd$_3B_4$O$_{12}$ in this study are compared with some inverse perovskites Sc$_3$InX (X = B, C, N), MAX phases and rock salt binary carbides [55, 56] those are exhibiting the sequence: $\sigma_p$ (selected inverse perovskites and MAX phases) $<\sigma_p$ (CaPd$_3B_4$O$_{12}$ double perovskites) $<<\sigma_p$ (binary carbides). Therefore, it is clear that dislocations can move in the selected MAX phases easily, but it is quite impossible in the case of the binary carbides. Since, $\sigma_p$ of CaPd$_3$B$_4$O$_{12}$ possessing between of MAX phases and binary carbides, dislocation can still move here, but not as easily as in MAX phases. However, dislocation movement in CaPd$_3$V$_4$O$_{12}$ may occur more easily for lower value of $\sigma_p$ in compare to CaPd$_3$Ti$_4$O$_{12}$.

**Table 2.** The elastic constants, $C_{ij}$ (GPa), bulk modulus, $B$ (GPa), shear modulus, $G$ (GPa), Young's modulus, $Y$ (GPa), Pugh's ratio, $G/B$, Poisson's ratio, $v$, Vickers hardness, $H_v$ (GPa), Cauchy pressure (GPa), universal anisotropic index $A^U$, Burger's vector $b$ (Å), interlayer distance $d$ (Å) and Peierls stress $\sigma_P$ (GPa) of CaPd$_3$B$_4$O$_{12}$ compounds.

| Compounds | $C_{11}$ | $C_{12}$ | $C_{44}$ | $B$ | $G$ | $Y$ | $G/B$ | $v$ | $H_v$ | Cauchy pressure | $A^U$ | $b$ | $d$ | $\sigma_P$ |
|---|---|---|---|---|---|---|---|---|---|---|---|---|---|---|
| CaPd$_3$Ti$_4$O$_{12}$ | 310 | 210 | 141 | 243 | 93 | 248 | 0.43 | 0.329 | 7.96 | 61 | 0.15 | 7.59 | 3.79 | 1.51 |
| CaPd$_3$V$_4$O$_{12}$ | 404 | 160 | 82 | 242 | 96 | 255 | 0.39 | 0.32 | 6.60 | 78 | 0.19 | 7.39 | 3.69 | 1.39 |

### 3.3. Electronic Properties

3.3.1. Band structure and Density of states

The electronic band structure, partial density of state (PDOS) and total density of state (TDOS) of CaPd$_3$B$_4$O$_{12}$ are depicted in Figs. 2 and 3, respectively. The electronic band structure confirms either a material is conductor or insulator. The PDOS and TDOS are defined in terms of number of states at occupied or unoccupied energy level in statistical and solid–state physics [57]. It indicates hybridization among orbital electrons as well as bonding characteristics within the compound. The Fermi level, $E_F$ is indicated by horizontal dashed line. In the band diagram, the purely valence and conduction bands are shown by magenta and blue lines respectively, whereas the bands crossing the $E_F$ indicated with orange lines. As can be seen from the band diagram of Fig. 2(a), there is no overlap between valance and conduction bands and exhibits direct band gap, $E_g$ ~0.46 eV for the CPTO using GGA-PBE potential. Since, GGA-PBE approximation underestimates the band gap of materials, thus we also employed TB-mBJ potential that predicted the transport properties of solids properly [58−62] to obtain the actual band gap of CPTO in this study. We found a band gap, $E_g$ ~0.88 eV using this potential as displayed in Fig. (2), these results suggest that the band gap of CPTO can be around this value. From the partial density of states of CPTO, we see the strong hybridization between Pd–4$d$ and O–2$p$ orbital electrons in the energy range of approximately −3.0 eV to $E_F$ for both potentials as presented in Fig. 3a. The band originates from the contribution of O–2$p$ orbital electrons in this energy range which is higher than Ca–4$p$/4$s$ and Ti–3$p$/3$d$ orbital electrons. Notably, in the midst of all orbitals Pd–4$d$ state near the $E_F$ contributes predominantly in the PDOS and hence TDOS for CaPd$_3$Ti$_4$O$_{12}$. Importantly, the valence bands adjacent to $E_F$ are smoother in contrast to conduction bands for this material using both potentials. The flatness of these bands originates from the robust hybridization of Pd–4$d$ and O–2$p$ states which is also reflected in the PDOS [28] of Fig. 3. The total DOS at $E_F$ for this compound is approximately 3.77 states/eV/unit cells is an indication of conductive nature of it as well. Moreover, the calculated electronic band structure and density of states (DOS) along the high symmetry directions of CaPd$_3$V$_4$O$_{12}$ are displayed in Fig. 4 using GGA approximation. From Fig. 4 it is clear that several valence and conduction bands cross the $E_F$ with large dispersion and overlap to each other is a strong evidence of metallic behavior of the double perovskite material which is nearly analogous to the previously reported articles [20, 21, 63, 64] as well. On the other hand, from Fig. 4 of right side we see the strong hybridization among Pd–4$d$, V–3$d$ and O–2$p$ orbital electrons in the energy range of −3.0 eV to $E_F$. At the Fermi level the most dominant contribution comes from V–3$d$ orbital electrons in compare to Pd–4$d$ and O–2$p$. Herein, the contribution of Ca–4$s$/4$p$ states is remarkable at $E_F$ but exhibits a small value than other orbitals. However, the individual contribution of O–2$p$ orbital electrons in the vicinity of $E_F$ is very similar to the common characteristics of PDOS of reported double and simple perovskite compounds [7, 10, 12, 13, 21]. Noteworthy, the strong hybridization among Pd–4$d$, V–3$d$ and O–2$p$ states imply a strong ternary ionic/covalent Pd–V–O bond in CaPd$_3$V$_4$O$_{12}$ perovskite. Therefore, the electronic structure calculations of CPVO demonstrate that the A′-site ions keep an important role for the metallic conduction of V−based perovskites [11].

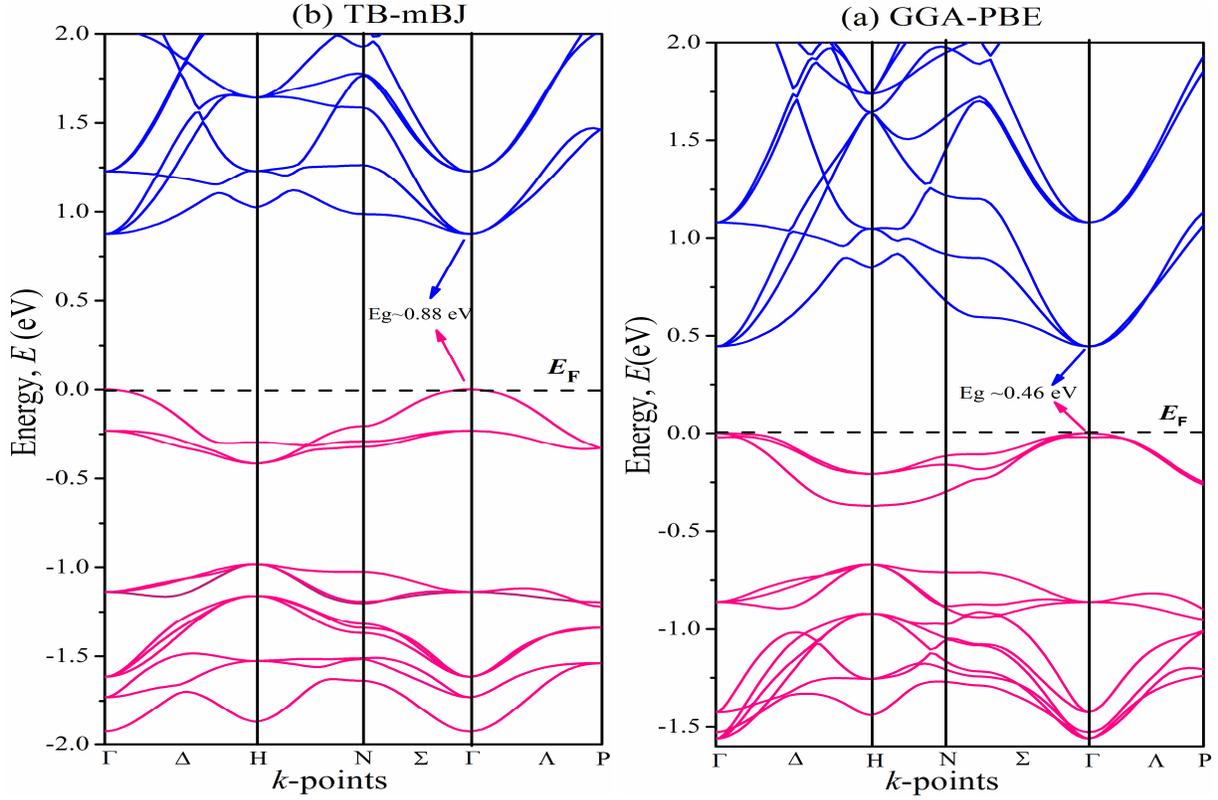

**Figure 2.** (Color online) Calculated band structures of $CaPd_3Ti_4O_{12}$ using (a) GGA–PBE and (b) TB–mBJ potentials along the high symmetry directions in the Brillouin zone at ambient conditions.

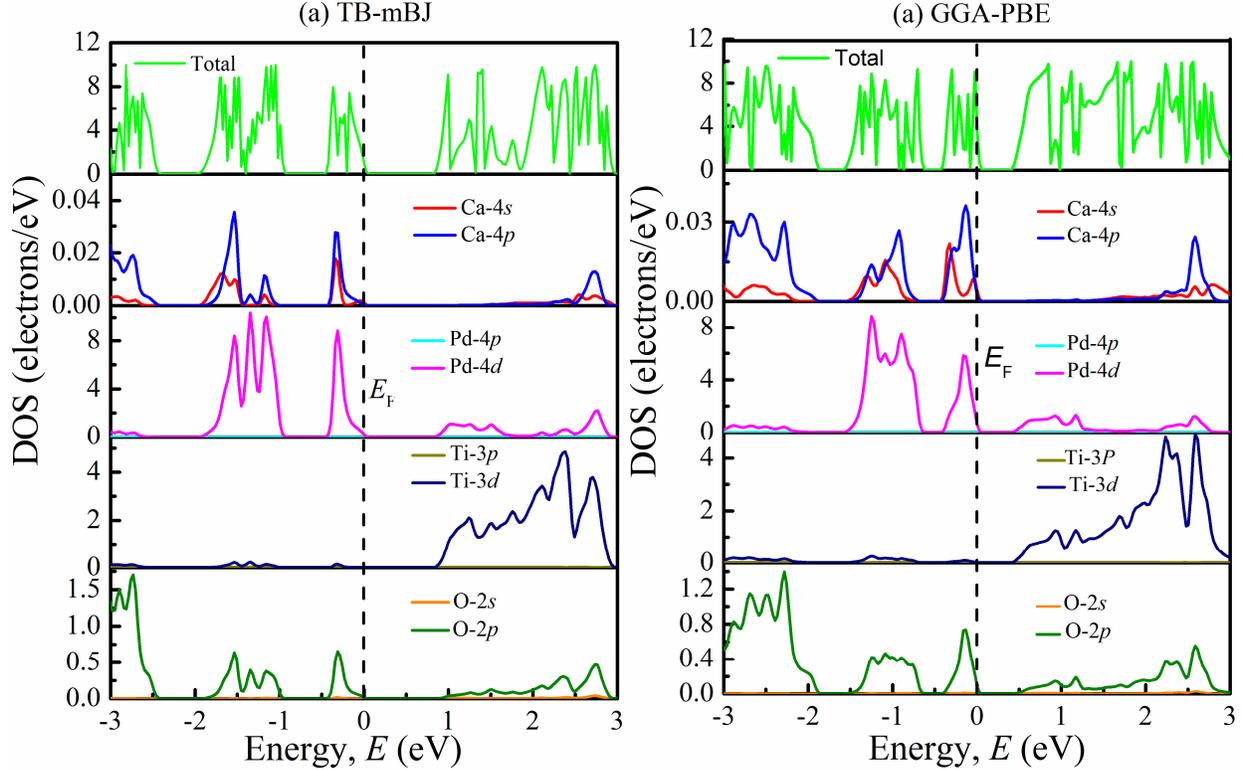

**Figure 3.** (Color online) total and partial electron energy density of states of $CaPd_3Ti_4O_{12}$ (a) GGA–PBE and (b) TB–mBJ potentials.

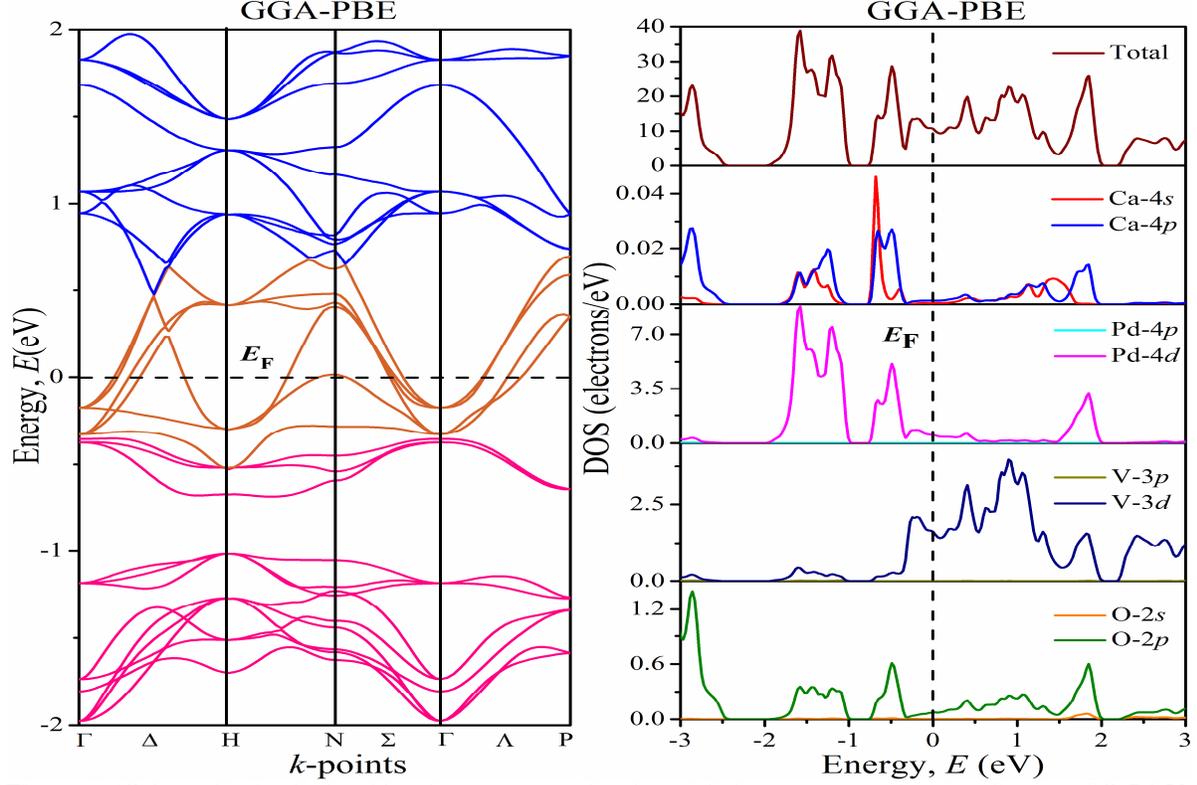

**Figure 4.** (Color online) calculated band structure, total and partial electron energy density of states of CaPd$_3$V$_4$O$_{12}$.

3.3.2. Electronic charge density and Fermi surface

The behavior of chemical bonding within a compound is analyzed by investigating the total electronic charge density map. Figure 5 presents the valence electronic charge density (in the units of e/Å$^3$) map of CaPd$_3$B$_4$O$_{12}$ double perovskite along (100) plane. The adjacent scale in the right side of both contour plots of Fig. 5 reveals the intensity of charge (electron) density. The blue and red colors in the scale indicate the light and high density of electrons, respectively. The Pd–O bond in CaPd$_3$Ti$_4$O$_{12}$ coincides with the hybridization between Pd–4$d$ and O–2$p$ orbitals as seen from Fig. 3, whereas Pd–V–O in CaPd$_3$V$_4$O$_{12}$ coincides with the strong hybridization among Pd–4$d$, O–2$p$ and V–4$d$ orbitals is evident from DOS in Fig. 4. From crystal structure it is also assumed that (Ti/V = $B$)O$_6$ octahedra built through ionic bonding. We see that O and Pd ions reveal ionic characteristics even though the charge density contour of oxygen is not entirely spherical but it shows ionic bonding. Further, the charge density distribution around all the atoms (excluding oxygen) is likely spherical which shows ionic nature and evident from Pd–O and Pd–V–O bonds of CPVO. This ionic nature is a consequence of metallic characteristics [65] as well though the band diagram of CPTO manifests the characteristics of a degenerate semiconductor. Therefore, a combination of chemical bonding namely ionic and metallic interactions exists within CaPd$_3$B$_4$O$_{12}$. Since, the O atom has higher electro negativity in contrast to others thus it is seen that the charge accumulates dominantly near O atom. The calculated electronic charge density distributions in different planes produced the same results is an indication of their isotropic nature. Furthermore, the calculated bond valence sum (BVS) of CPTO indicated the ionic model as Ca$^{2+}$Pd$^{2+}_3$Ti$^{4+}_4$O$_{12}$ whereas that for CPVO implies an ionic model of Ca$^{2+}$Pd$^{2+}_3$V$^{4+}_4$O$_{12}$. Since, the charge density distribution is essentially spherical around all the atoms for compounds, therefore showing ionic nature of Pd–O and Pd–V–O bonds.

For being a diamagnetic insulator/semiconductor we could not calculate Fermi surface of CaPd$_3$Ti$_4$O$_{12}$. On the other hand, we calculate the Fermi surface of CaPd$_3$V$_4$O$_{12}$ for the bands crossing $E_F$ is shown in Fig. 6. It is noticed from the figure that an electron–like sheet with cubic cross section is centered along the Γ–Z direction of the Brillouin zone. There is also a hole–like Fermi surface around X–point. At the corner of the Brillouin zone, another electron–like Fermi surface around R–point is connected with the hole like Fermi surface. Therefore, it is evident that both electron and hole–like Fermi surfaces are present in CPVO which indicates the multiple–band nature of this compound.

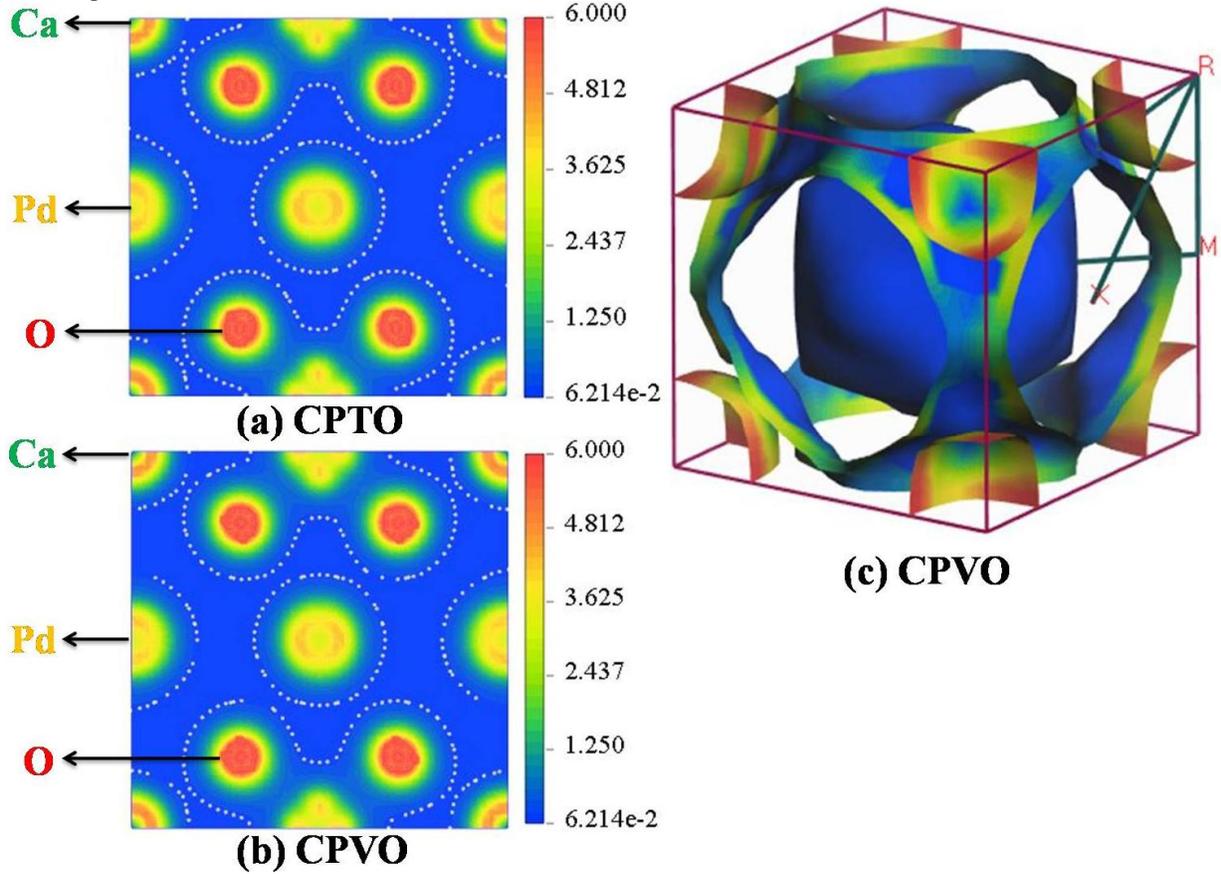

**Figure 5.** Electronic charge density of (a) $CaPd_3Ti_4O_{12}$ and (b) $CaPd_3V_4O_{12}$. Fermi surface topology of $CaPd_3V_4O_{12}$ double perovskite is depicted in (c).

**3.4. Optical Properties**

The optical properties of perovskite materials play a significant role in the applications of optoelectronics and nanoelectronics devices [66] and are very effective to explain electronic structure of solids. The various optical parameters such as dielectric function, refractive index, photo−conductivity, absorption, reflectivity and loss function have been calculated to investigate electronic structure and optical applications of $CaPd_3B_4O_{12}$. The real ($\varepsilon_1$) and imaginary ($\varepsilon_2$) parts of the dielectric function precisely describe the optical behaviors of materials [67] and has been studied by the frequency-dependent dielectric function $\varepsilon(\omega) = \varepsilon_1(\omega) + i\varepsilon_2(\omega)$ which has a direct relation to the electronic configurations of materials. Figures 7a and 7b depict the spectra of real and imaginary parts of dielectric function in respect to photon energy for $CaPd_3B_4O_{12}$. The real part of the dielectric constant in the zero frequency limits is called the electronic part of the static dielectric constant. So the static dielectric constant of $CaPd_3Ti_4O_{12}$ is 11.07 (Fig. 7a). It is observed that the real part ($\varepsilon_1$) goes to below from zero at around 8.17 eV and back to zero at about 12.18 eV. Notably, the negative value of the $\varepsilon_1$ implies the Drude−like behaviour of $CaPd_3Ti_4O_{12}$. The imaginary part $\varepsilon_2$ approaches to zero around 40.64 eV. Besides, $CaPd_3V_4O_{12}$ is a metallic material that is clearly seen from the band diagram (Fig. 4). Owing to metallic nature of CPVO, we employ the Drude plasma frequency of 3 eV and damping factor of 0.05 eV to investigate dielectric properties. The peak of the real part is connected to the electron excitation and is mainly arisen due to the intraband transitions. It is well known for metal and metal−like systems that the intraband contributions come from the conduction electrons mainly in the low−energy part of the spectra. So, we may conclude that the intraband contributions in $CaPd_3V_4O_{12}$ arise from the conduction electrons. It is significant that the $\varepsilon_2(\omega)$ approaches zero at around 30 eV in the ultraviolet energy region, by which we can predict that the material is transparent and optically anisotropic as well. The anisotropic nature of $CaPd_3V_4O_{12}$ was also revealed by the elastic properties.

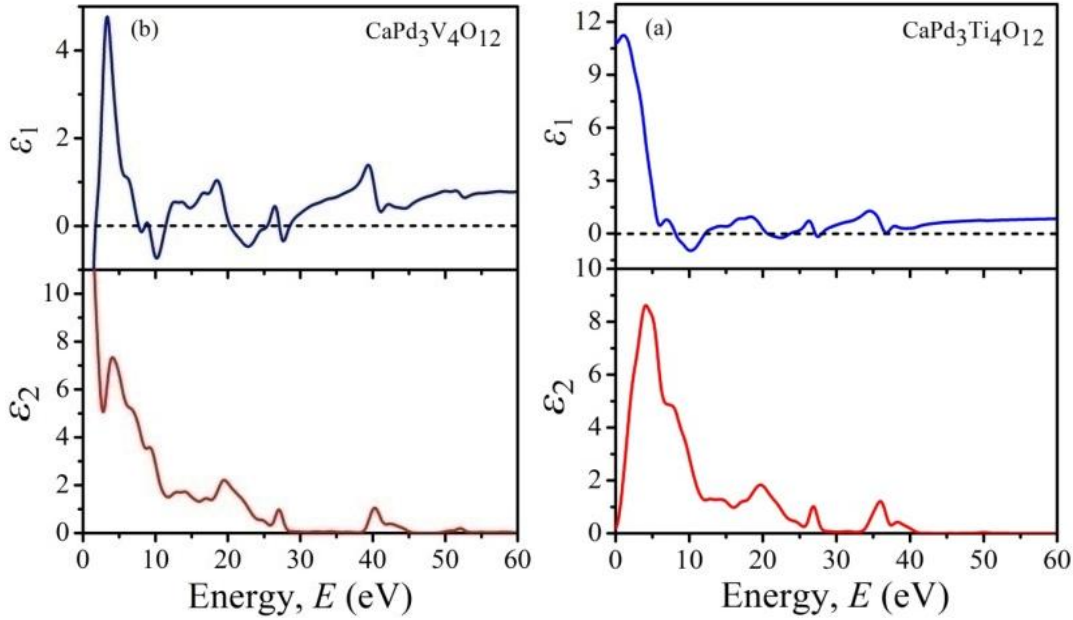

**Figure 7.** Energy dependent dielectric function (real part, $\epsilon_1$ and imaginary part, $\epsilon_2$) of (a) $CaPd_3Ti_4O_{12}$ and (b) $CaPd_3V_4O_{12}$.

A material to be used in photovoltaic system high optical conductivity, high absorption, and high value of refractive index and less emissivity are required conditions. Among these, the optical compatibility of a material is usually estimate based on the sense of refractive index to be applied in optical devices including waveguides, data storage media, photonic crystals and so on [68]. The refractive index ($n$) implies the phase velocity and the extinction coefficient ($k$) and the amount of absorption loss, when the electromagnetic wave (as light) travels through the material. The refractive index ($n$) and the extinction coefficient ($k$) of $CaPd_3B_4O_{12}$ are shown in Figs. 8.

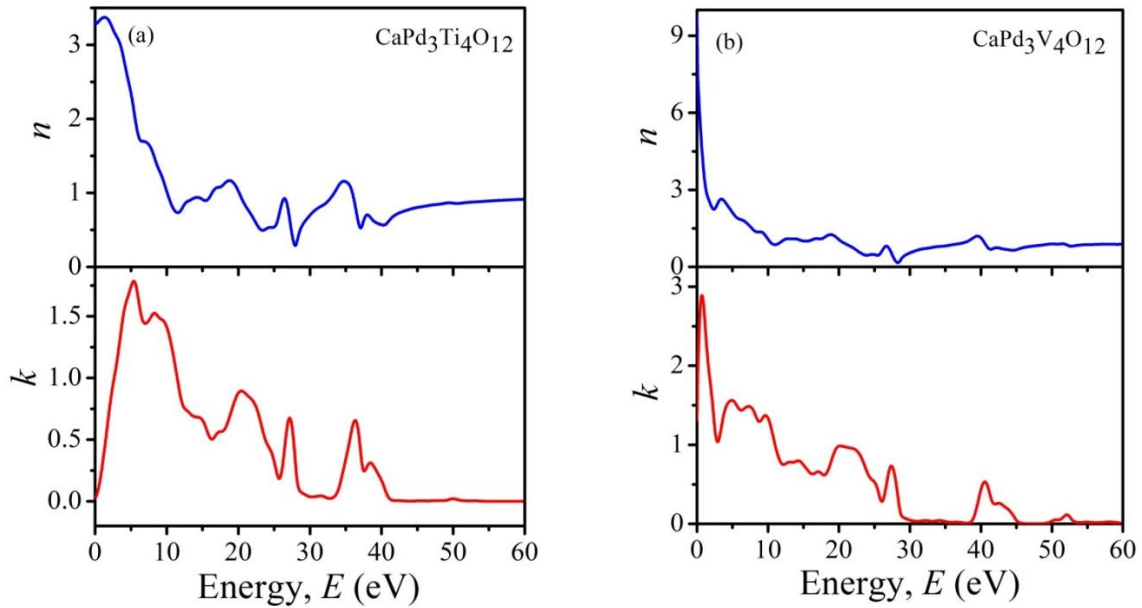

**Figure 8.** (Color online) The refractive index ($n$) and the extinction coefficient ($k$) of (a) $CaPd_3Ti_4O_{12}$ and (b) $CaPd_3V_4O_{12}$.

The calculated values of the static refractive index $n(0)$ of CPTO and CPVO are 3.32 (Fig. 8a) and 9.78 (Fig. 8b), respectively, and decreases in the high energy region as well. The calculated value of refractive index $n(0)$ of

CaPd$_3$Ti$_4$O$_{12}$ is little higher than that of K$_2$Cu$_2$GeS$_4$ semiconductor but in the vicinity of GaAs semiconductor [28]. Figure 9(a) shows the real part of conductivity spectrums of CaPd$_3$B$_4$O$_{12}$. It is evident that the photoconductivity of CaPd$_3$Ti$_4$O$_{12}$ has not initiated at zero photon energy because of narrow band gap for the material which can also be seen from the electronic band diagram. On the other hand, the photoconductivity of CaPd$_3$V$_4$O$_{12}$ compound starts from zero photon energy owing to its metallic characteristics. However, the maximum photo conductivities are 5.25 and 5.23 found at 35.96 and 19.58 eV for CaPd$_3$Ti$_4$O$_{12}$ and CaPd$_3$V$_4$O$_{12}$, respectively.

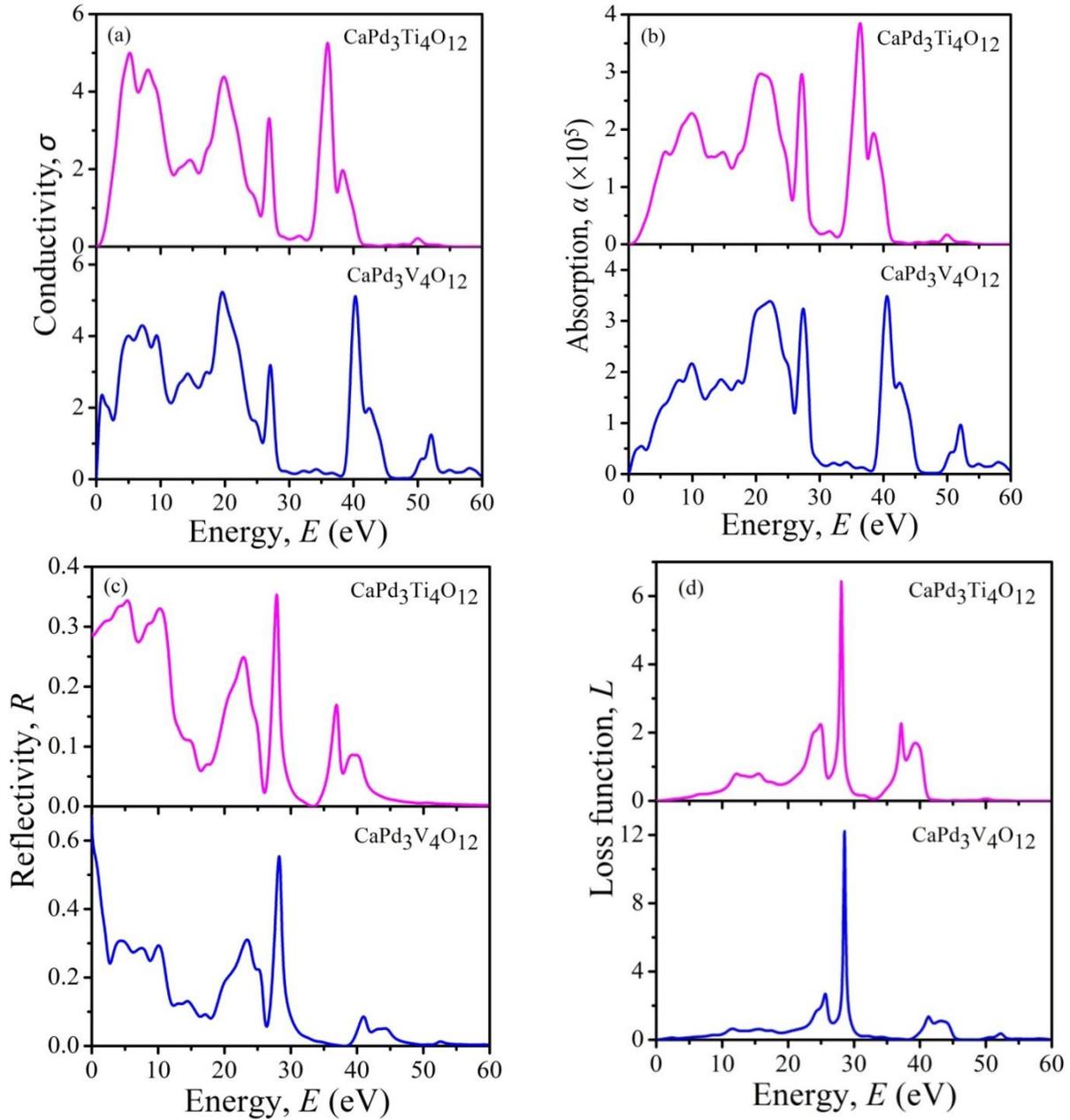

**Figure 9.** (Color online) Energy dependent of (a) real part of conductivity ($\sigma$), (b) absorption ($\alpha$), (c) reflectivity ($R$), and (d) loss function ($L$) of CaPd$_3$B$_4$O$_{12}$.

It is noteworthy, the photoconductivity and thus electrical conductivity for a semiconductor enhances with absorbed photons [69]. Fig. 9(b) represents the absorption coefficient data and spectrums of CaPd$_3$B$_4$O$_{12}$ that are related to the optimal solar energy conversion efficiency which implies the penetration capability of materials before being absorbed. It is visualized that the absorption for CaPd$_3$Ti$_4$O$_{12}$ begins ascending at ~0.90 eV owing to the semiconducting behavior together with calculated band gap of 0.88 eV which is acquainted as absorption edges as well. Whereas, the absorption of $CaPd_3V_4O_{12}$ metal starts increasing from zero energy because of overlapping of bands (metallicity) as seen from band structure calculation. However, generally the intraband contribution mainly affects in the low energy infrared part of the spectra. In addition, the peaks in the high energy region of the absorption and conductivity spectra may arise because of the interband transition. Therefore, it is very clear from Figs. 9(a) and (b) that the variation of the conductivity spectra is almost similar to the absorption spectra for both these compounds. Hence, the photoconductivity of $CaPd_3B_4O_{12}$ increases as a result of absorbing photons [70]. The reflectivity spectrum is a crucial optical parameter to determine all the optical constants using Kramers-Kroning relations and the reflectivity spectra for $CaPd_3B_4O_{12}$ are depicted in Fig. 9(c). It is observed that the reflectivity spectra in case of both compounds have initiated from the zero frequency is regarded as the static portion of reflectivity. Interestingly, a moderate reflectivity of approximately 28% for $CaPd_3Ti_4O_{12}$ but a very high reflectivity of ~66% is seen for $CaPd_3V_4O_{12}$ in the infrared region. Moreover, some peaks are found for both materials in the high energy region due to interband transition. The photon energy loss spectrums (L) of $CaPd_3B_4O_{12}$ are shown in Fig. 9(d). Loss function is a significant parameter which reveals the amount of energy loss of a fast electron when it passes through a material [71]. In the graph, the peaks are related with the plasma resonance and its associated frequency is called the plasma frequency $\omega_p$ [72]. The highest peaks are found at 28.07 and 28.57 eV for $CaPd_3Ti_4O_{12}$ and $CaPd_3V_4O_{12}$ respectively, which reveal the plasma frequency of these two compounds.

### 3.5. Population Analysis

Many interesting information are obtained from Mulliken atomic population analysis which are needed to understand the chemical bonding nature in solids [73]. Mulliken orbitals and Mulliken atomic populations obtained in this work as well as the estimated overlap populations for nearest neighbour atoms of $CaPd_3B_4O_{12}$ are listed in Table 3 and Table 4, respectively. If the bond overlap population is zero, the bond is perfectly ionic bond, whereas the value greater than zero indicates the increasing levels of covalency [74].

**Table 3.** Mulliken atomic population analysis of $CaPd_3B_4O_{12}$ compounds.

| Compounds | Species | Mulliken atomic populations | | | | Charge (e) |
|---|---|---|---|---|---|---|
| | | s | P | D | Total | |
| $CaPd_3Ti_4O_{12}$ | Ca | 2.10 | 5.99 | 0.59 | 8.67 | 1.33 |
| | Pd | 2.53 | 6.17 | 8.77 | 17.47 | 0.53 |
| | Ti | 2.28 | 6.49 | 2.14 | 10.91 | 1.09 |
| | O | 1.83 | 4.77 | 0.00 | 6.60 | -0.60 |
| $CaPd_3V_4O_{12}$ | Ca | 2.10 | 5.99 | 0.60 | 8.70 | 1.30 |
| | Pd | 2.50 | 6.13 | 8.82 | 17.45 | 0.55 |
| | V | 2.30 | 4.48 | 3.24 | 12.03 | 0.97 |
| | O | 1.83 | 4.74 | 0.00 | 6.57 | -0.57 |

From Table 4, the atomic bond populations for both compounds are positive and negative. Negative value indicates ionic nature but the value greater than zero reveals the covalent nature of these compounds. Therefore, we can conclude that both ionic and covalent bond exist within $CaPd_3B_4O_{12}$ which agree with the electronic charge density map. It is noted from Table 4 that O–V bond in $CaPd_3V_4O_{12}$ is more covalent than other bonds in $CaPd_3B_4O_{12}$.

**Table 4.** Calculated Mulliken bond number $n^\mu$, bond length $d^\mu$ and bond overlap population $P^\mu$, of $CaPd_3B_4O_{12}$ compounds.

| Compounds | Bonds | $n^\mu$ | $d^\mu$(Å) | $P^\mu$ |
|---|---|---|---|---|
| $CaPd_3Ti_4O_{12}$ | O–Pd (I) | 24 | 2.02280 | 0.27 |
| | O-Pd (II) | 24 | 2.85409 | -0.20 |
| $CaPd_3V_4O_{12}$ | O–Pd (I) | 24 | 2.04406 | 0.25 |
| | O-Pd (II) | 24 | 2.76528 | -0.19 |
| | O–V (III) | 48 | 1.91992 | 0.49 |

## 3.6. Thermodynamic properties

Various thermodynamic properties such as Debye temperature ($\Theta_D$), melting temperature ($T_m$) and thermal conductivity of the CaPd$_3B_4$O$_{12v}$ perovskites are calculated to understand their behaviour under high temperatures and high pressures. Debye temperature ($\Theta_D$) plays an important role to address some interesting physical properties such as lattice vibrations, thermal conductivity, melting point, specific heat and so on. Interestingly, $\Theta_D$ is directly related with the lattice thermal conductivity to evaluate thermoelectric performance of a material. Thus, it is noteworthy to calculate Debye temperature of these perovskite to have an idea respecting thermal conductivity which can be calculated by using the following expression [75];

$$\theta_D = \frac{h}{k_B}\left[\frac{3m}{4\pi}\left(\frac{N_A\rho}{M}\right)\right]^{\frac{1}{3}} v_m$$

where, $h$ and $k_B$ are the Planck's and Boltzmann constants, respectively. $V$ is the volume of unit cell, $n$ is the number of atoms within a unit cell, and $v_m$ is the average sound velocity. The parameter $v_m$ is calculated by the following equation,

$$v_m = \left[\frac{1}{3}\left(\frac{2}{v_t^3} + \frac{1}{v_l^3}\right)\right]^{-\frac{1}{3}}$$

Here, $v_l$ and $v_t$ are the longitudinal and transverse sound velocities, respectively. By using the value of bulk modulus, $B$ and shear modulus, $G$, the $v_l$ and $v_t$ can be estimated using the following expressions,

$$v_l = \left(\frac{B+\frac{4}{3}G}{\rho}\right) \text{ and } v_t = [G/\rho]^{1/2}.$$

The melting temperature of a crystal is an essential entity for the application in a heating system that can be calculated employing the expression [76] as follows,

$$T_m = \left[553\,K + \left(\frac{5.91\,K}{GPa}\right)C_{11}\right] \pm 300\,K$$

In cubic structure, the axial lengths are equal; therefore, the elastic constants $C_{11}$, $C_{22}$ and $C_{33}$ are also equal. The calculated melting temperature ($T_m$) of CPVO and CPTO materials is high (Table 5) which makes them favorable for elevated temperature applications. All the calculated values of $\Theta_D$, $T_m$, $v_m$, $v_t$ and $v_l$ for the titled compounds are listed in Table 5. From Table 2&5 it is seen that the relatively high value of $\Theta_D$, $T_m$ and elastic constants (B, G and Y) imply the hardness of the perovskite.

**Table 5.** The Calculated density ($\rho$), longitudinal, transverse, and average sound velocities ($v_l$, $v_t$, and $v_m$) Debye temperature ($\Theta_D$) and melting temperature ($T_m$) double perovskites CaPd$_3B_4$O$_{12}$.

| Compounds | $\rho$ (g/cm$^3$) | $v_l$(km/s) | $v_t$(km/s) | $v_m$(km/s) | $\Theta_D$(K) | $T_m$(K) |
|---|---|---|---|---|---|---|
| CaPd$_3$Ti$_4$O$_{12}$ | 5.935 | 7.928 | 3.99 | 4.479 | 608 | 2385.1 ± 300 |
| CaPd$_3$V$_4$O$_{12}$ | 6.343 | 7.637 | 3.890 | 4.358 | 604 | 2940.7 ± 300 |

## 3.7. Thermoelectric transport properties of CaPd$_3$Ti$_4$O$_{12}$

The flat and narrow band gap semiconductors are potential candidates for thermoelectric device application. The electronic band structure of the studied compounds reveal that CPVO is metallic and CPTO is semiconducting. The narrow band gap of CPTO inspired us to study its thermoelectric transport properties. Recently, thermoelectric properties of numerous semiconductors have been investigated based on DFT calculations [58–62]. The temperature dependence thermoelectric transport properties of CPTO are calculated using GGA-PBE and TB-mBJ potentials are presented in Figs. 10. For a potential thermoelectric material, higher Seebeck coefficient and electrical conductivity are expected for higher power factor but their coexistence is quite rare. The calculated Seebeck coefficient decreases with temperature for both potentials. For GGA-PBE potential the calculated Seebeck coefficient is lower than that obtained using TB-mBJ potential. This is expected because GGA-PBE potential underestimates the band gap. The Seebeck coefficient at 300 K for GGA-PBE and TB-mBJ potentials are 204 and 233 mV/K, respectively. Fig. 10 (b) and Fig. 10 (d) represent the temperature dependence of electrical conductivity and electronic thermal conductivity and we notice that both are increasing with temperature. As the temperature increases, the carrier concentration increases and hence electrical conductivity and electronic thermal conductivity both increases. This behavior implies the semiconducting nature of CPTO. The maximum power factor 11.9 mWcm$^{-1}$K$^{-2}$ (with $\tau = 10^{-14}$ s) using TB-mBJ potential obtained at 800 K as shown in Fig. 10 (c) is slightly higher than that of SnSe, a promising thermoelectric material [77]. The dimensionless figure of merit is predicted in Fig. 10 (e) and it is observed that ZT ~0.8 reaches maximum in the temperature range 800 to 1100 K. However, in the investigation we did not calculate lattice thermal conductivity. The lattice thermal conductivity is directly related with lattice vibration (phonon scattering). In general, at low temperature due to low phonon scattering lattice thermal conductivity is much higher compared to electronic thermal conductivity but at higher temperature it decreases rapidly and electronic thermal conductivity is dominating in semiconducting materials. Therefore, CPTO is a promising material for thermoelectric energy conversion and our predicted results will inspire experimentalists to measure thermoelectric properties for practical use.

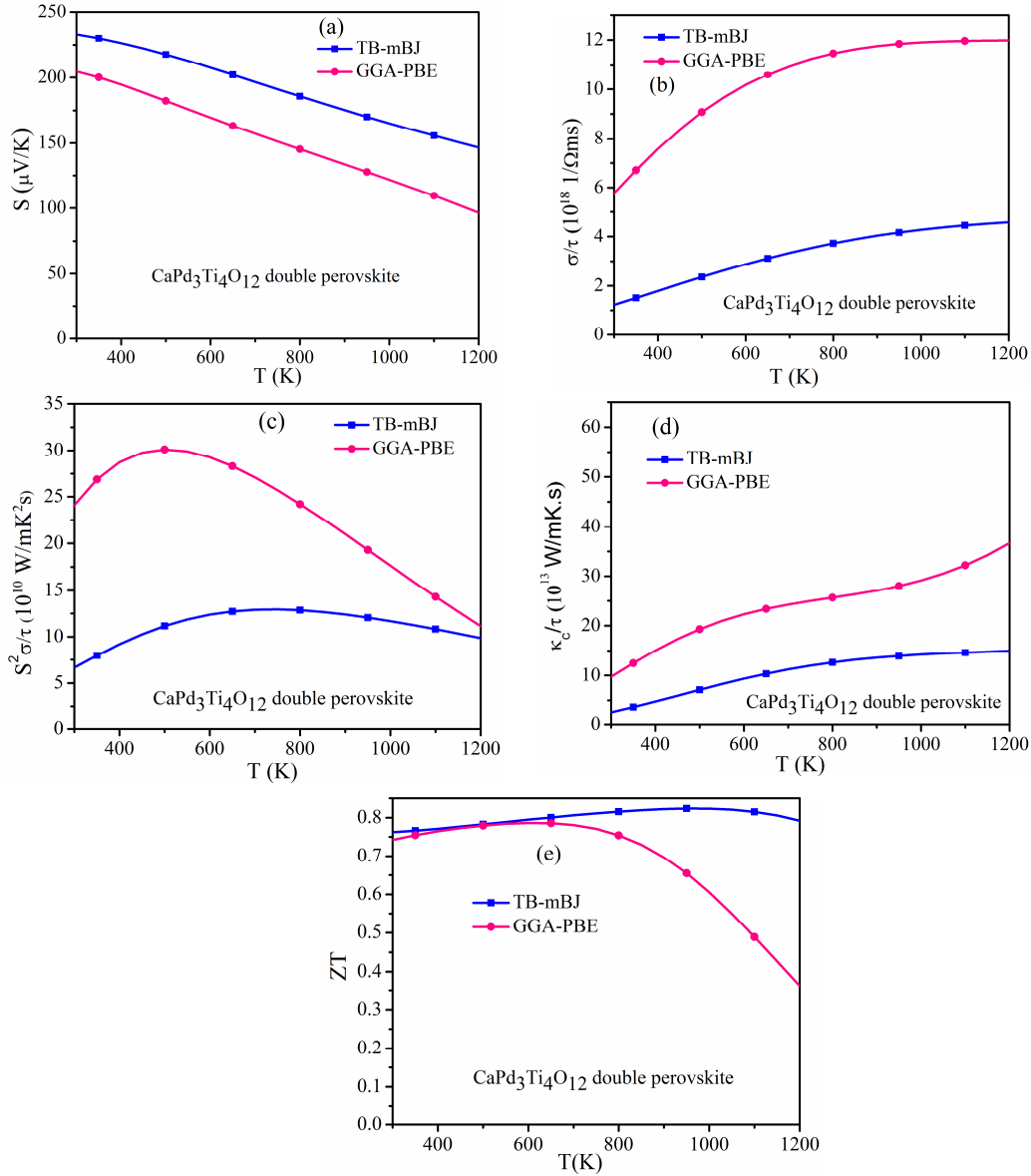

**Figure 10.** Thermoelectric transport properties as a function of temperature of CaPd$_3$Ti$_4$O$_{12}$ double perovskite using both GGA-PBE and TB-mBJ potentials (a) Seebeck coefficient, (b) Electrical conductivity, (c) Power factor, (d) Electronic thermal conductivity and (e) Dimensionless figure of merit.

## 4. Conclusions

In this paper, we have performed DFT calculations on structural, mechanical, electronic, optical, thermoelectric and thermodynamic properties of synthesized cubic perovskites $CaPd_3Ti_4O_{12}$ and $CaPd_3V_4O_{12}$. The calculated lattice parameters of these compounds are in good agreement with the available experimental data that clarify the reliability of calculation. The calculated elastic constants also satisfy the mechanical stability conditions for both compounds. The computed polycrystalline elastic constants and Universal anisotropy index imply the hardness and elastically anisotropy of $CaPd_3B_4O_{12}$ ($B$ = Ti, V). Moreover, the Poisson's ratio, Pugh's ratio and Cauchy pressure reveal the ductile nature of them. The calculated band structures of $CaPd_3Ti_4O_{12}$ exhibit a direct band gap of ~0.88 eV in contrast to the metallic nature of $CaPd_3V_4O_{12}$. The bonding properties reveal the coexistence of ionic and covalent bond within $CaPd_3B_4O_{12}$ which has been demonstrated from electronic charge density maps and bonding population analysis. Furthermore, the multiple−band nature of $CaPd_3V_4O_{12}$ compounds is found evident from the existence of both electron and hole−like Fermi surfaces. The static dielectric constants of 11.07 ($n(0)$ = 3.23)) and 5.78 ($n(0)$ = 9.78)) was observed for $CaPd_3Ti_4O_{12}$ and $CaPd_3V_4O_{12}$ respectively, from the computed dielectric function. Moreover, the high value of reflectivity of these compounds suggests that the materials could be used as promising coating material to attenuate solar heating. Very similar variational photoconductivity and absorption spectra are visualized for the both double perovskites and their conductivity increase with absorbing photons. However, the calculated optical properties of them display slight anisotropy with respect to used photon energy. Moreover, various thermodynamic properties such as $\Theta_D$, $T_m$, $v_m$, $v_t$ and $v_l$ have been calculated among which Debye temperature and melting temperature exhibit reasonably high values. Generally, a higher Debye temperature comprise in a higher phonon thermal conductivity and versa−vice. The Seebeck coefficient gradually decreases with the increase of temperature. The electronic conductivity and electronic thermal conductivity both are directly related with carrier concentration and increases with the increase of temperature. This behavior indicates the semiconducting nature of $CaPd_3Ti_4O_{12}$. The calculated power factor increases gradually with the increase of temperature and then decreases again after 800 K. The maximum power factor ~11.9 $mWcm^{-1}K^{-2}$ with t = $10^{-14}$ s was found at 800 K for the CPTO semiconductor. The predicted thermoelectric figure of merit (ZT = 0.8) at 800 K is very close to unity and making the studied material as auspicious for thermoelectric device application. However, the reduction of thermal conductivity or enhancement of power factor can be achieved by nano-structuring or doping with suitable element in CPTO system. We hope our predicted results will inspire experimentalists for the measurement of thermoelectric properties in CPTO and further improvement in ZT by nano-structuring or doping for practical use in converting waste heat to electricity.